\documentclass[pra,amsmath,twocolumn,showpacs]{revtex4}
\usepackage{graphicx}
\usepackage{hyperref}
\usepackage[T1]{fontenc}

\newcommand{\eeqref}[1]{Eq.~(\ref{#1})}
\newcommand{\avg}[1]{\langle #1\rangle}
\begin{document}
\date{\today}
\title{Pairwise entanglement and readout of atomic-ensemble and optical wave-packet modes in traveling-wave Raman interactions}
\author{Wojciech Wasilewski}
\affiliation{Institute of Physics, Nicolaus Copernicus University, Grudzi{\k a}dzka 5, 87-100 Toru{\'n}, Poland}
\author{M. G. Raymer}
\affiliation{Department of Physics and Oregon Center for Optics, University of Oregon, Eugene, Oregon 97403}

\begin{abstract}
We analyze  quantum entanglement of Stokes light and atomic electronic polarization excited during single-pass,
linear-regime, stimulated Raman scattering in terms of optical  wave-packet modes and atomic-ensemble spatial modes.
The output of this process is confirmed to be decomposable into multiple discrete, bosonic mode pairs, each pair
undergoing independent evolution into a two-mode squeezed state. For this we extend the Bloch-Messiah reduction
theorem,  previously known for discrete linear systems (S.~L. Braunstein,  {\em Phys. Rev. A},  vol.~71, p.~055801,
2005). We present typical mode functions in the case of one-dimensional scattering in an atomic vapor. We find that in
the absence of dispersion, one mode pair dominates the process, leading to a simple interpretation of entanglement in
this continuous-variable system. However, many mode pairs are excited in the presence of dispersion-induced temporal
walkoff of the Stokes, as witnessed by the photon-count statistics. We also consider the readout of the stored atomic
polarization using the anti-Stokes scattering process. We prove that the readout process can also be decomposed into
multiple mode pairs, each pair undergoing independent evolution analogous to a beam-splitter transformation. We show
that this process can have unit efficiency under realistic experimental conditions. The shape of the output light wave
packet can be predicted. In case of unit readout efficiency it contains only excitations originating from a specified
atomic excitation mode.
\end{abstract}

\maketitle
\section{Introduction}
Efficient entanglement generation, distribution and storage are the basic requirements for development of successful
quantum information technologies, including proposals  for realizing quantum information networks
\cite{LukinRMP03,DuanNature01}. They comprise entanglement generation by Raman scattering, storage in collective
polarization of atomic ensembles and release by anti-Stokes Raman scattering. Such a network can be used, among other
things, for a quantum cryptography \cite{EkertPRL91} and quantum teleportation \cite{BennettPRL93}. Those protocols
deal with spin or polarization entanglement. Another important possibility is the entanglement of continuous variables,
such as light quadrature amplitudes or molecular vibrational coordinates. Stokes Raman scattering
\cite{Raymer1DTheory81,Raymer3DTheory85} generates entanglement between light and atoms, and this in turn can be used
for entangling distant atomic ensembles \cite{DuanPRL00,Julsgaard2001,RaymerJMO04}. Such an entanglement can be
exploited for continuous-variable quantum cryptography \cite{RalphPRA00} or teleportation \cite{BraunsteinPRL1998}. In
the Stokes or anti-Stokes configuration, Raman scattering can be used for transferring a quantum state of light to
atoms or vice versa \cite{KozehekinPRA00,Chou2004,EisamanPRL04,Felinto2005}. Already early works demonstrated phase
memory of atomic polarization induced by Raman scattering and its robustness against decoherence
\cite{BelsleyCoherence93}, thus a Raman medium offers the possibility to realize long-lived continuous-variable quantum
memory.

Parallel to Raman scattering techniques, the phenomenon of Electromagnetically Induced Transparency (EIT) is being
extensively studied. In this completely resonant situation, weak probe light can be slowed down and stopped in a form
of localized atomic polarization by controlling a strong pump \cite{FleischhauerPRL00,PhilipsPRL01,dantan:043810}. In
this form of a quantum memory \cite{FleischhauerPRA02}, the light bandwidth is narrower, compared to the off-resonant
Raman interaction, due to the resonant character of this phenomenon \cite{LukinRMP03}.

\begin{figure}[b]
  \center
  \includegraphics[scale=0.9]{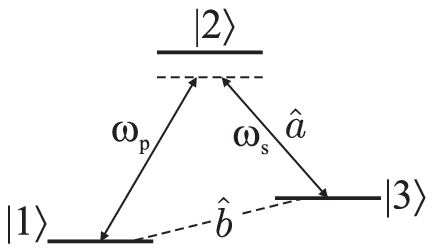}\hfill\includegraphics[scale=0.9]{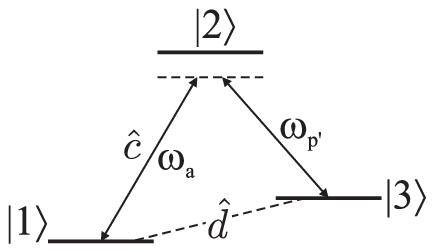}
  \caption{Relevant atomic levels in the Stokes and anti-Stokes (right panel) Raman scattering. Initially atoms reside in
  state $|1\rangle$. During the first scattering process (left panel) they are driven by strong pump field at a frequency $\omega_p$,
  which creates Stokes optical field $\hat a$ at a frequency $\omega_s$ and atomic polarization $\hat b$.
  Subsequently, the atoms are subjected to a strong field at a frequency $\omega_{p'}$ which erases atomic polarization
  we will denote by $\hat d$ shifting the excitation into anti-Stokes optical field $\hat c$ at a frequency $\omega_a$ (right panel).}
  \label{fig:levels}
\end{figure}

\begin{figure}
  \center
  \includegraphics[scale=0.75]{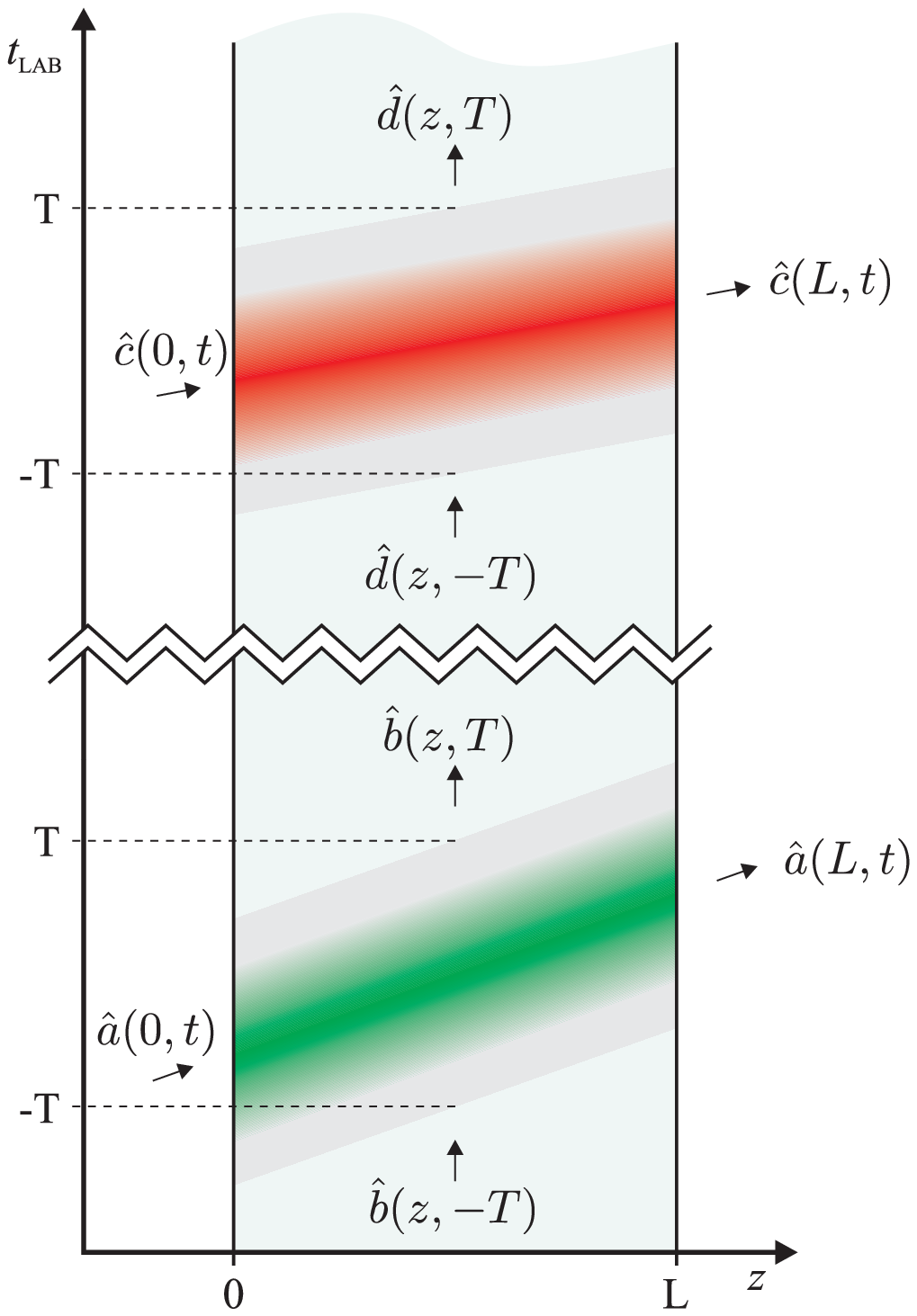} \caption{Pictorial representation of the Stokes and subsequent
  anti-Stokes scattering processes (see Sec.~\ref{sec:aStokes}) in the laboratory reference frame. The vertical stripe
  represents the cell in which the nonlinear medium is contained, the shaded parallelograms are the spatio-temporal
  interaction boundaries, the grey (green in color) stripe represents Stokes pump field envelope $A_p(z,t)$, while the
  dark grey (red) stripe represents anti-Stokes pump field envelope $A'_p(z,t)$. Letters with hats are the annihilation
  operators of the input and output quantum fields, defined at the respective boundaries of the interaction regions.}
  \label{fig:overall}
\end{figure}

In this paper we focus on the precise quantum description of the generation of entanglement between light and an atomic
ensemble by spontaneous and stimulated Stokes Raman scattering, followed by the readout of the atomic polarization
state onto an anti-Stokes field, as depicted in Fig~\ref{fig:levels} and \ref{fig:overall}. Following Raymer
\cite{RaymerJMO04}, we will use an input-output formalism to describe these interactions. As was already conjectured
there and proved approximately by numerical means, the Raman scattering process can be decomposed into a discrete set
of two-mode entanglement processes. Each process is described by a Bogoliubov two-mode squeezing transformation, which
linearly mixes photon annihilation and creation operators. Each elementary process squeezes the phase sum of excitation
amplitudes of optical and atomic-ensemble collective bosonic modes. Below we shall present a proof of this
decomposition, an extension of the Bloch-Messiah reduction \cite{BraunsteinBM}, and demonstrate that it is applicable
as long as the input-output relations are linear relations, involving mixing creation and annihilation operators.

We also consider the readout of the stored atomic polarization using the anti-Stokes scattering process. We prove that
the readout process can also be decomposed into multiple mode pairs, each pair undergoing independent evolution
analogous to a beam-splitter transformation. For this, we derive a canonical form of an all-linear unitary transforation
acting on a bipartite system.

Throughout this paper we  keep in mind the idea of an experiment with Raman scattering in a thermal vapor of $^{87}$Rb,
pumped by highly detuned, sub-nanosecond laser pulses. Our idealization comprises neglecting the influence of sidewards
spontaneous emission, as well as atomic polarization damping, atom loss and density fluctuations. We will also restrict
ourselves to a one-dimensional model, which is justified approximately in the case when the Fresnel number of the pump
beam equals unity or less \cite{Raymer3DTheory85}. However, we will include the effects of dispersion in the vapor,
which introduces appreciable group-velocity difference between the pump and scattered wave under realistic experimental
conditions. We find that dispersion significantly increases the number of independent modes excited in the Raman
scattering process, since the Stokes photons born in different parts of the medium  become distinguishable by their
delay respective to the pump pulse. Dispersion also can reduce the efficiency of the readout process.

Our work stems from early studies of stimulated Raman scattering \cite{Raymer1DTheory81,Raymer3DTheory85,RaymerPiO90}, and we adopt
the formalism of these works. We will also draw from early results on quantum energy
\cite{WalmsleyStatistics83,RaymerStatistics82} and phase \cite{KuoStatistics91} statistics of Stokes pulses.

In Sec.~\ref{sec:StokesTheory} we give theoretical foundations of the decomposition used to analyze the structure of
squeezing generated in the Stokes scattering process. The specific case of pulsed pump fields and nonzero dispersion,
which needs to be solved numerically, is addressed in Sec.~\ref{sec:StokesNumeric}. The consequences of the multimode
character of the output field for photon-count statistics are studied in Sec.~\ref{sec:PC}. The problem of reading out
of the atomic polarization by anti-Stokes scattering is discussed in Sec.~\ref{sec:aStokes}.

\section{Pair-wise Entanglement Operation}\label{sec:StokesTheory}
We begin developing our model by idealizing the process of stimulated Raman scattering as a three-wave mixing and
restricting ourselves to one dimension \cite{Raymer1DTheory81}. This is justified provided that atomic saturation and
pump field depletion can be neglected while the pump Rayleigh range is comparable to or shorter than the medium length
\cite{Raymer3DTheory85}.

We formulate the equations in the moving reference frame of the Stokes wave
\begin{equation}\label{eq:tStokes}
t=t_\text{LAB}-\frac{z-L/2}{v_\text{gr,s}}
\end{equation}
where $t_\text{LAB}$ is the time in the laboratory frame, $v_\text{gr,s}$ is the Stokes group velocity, while $L$ is
the medium length. According to the above expression, the laboratory and Stokes reference frames meet in the middle of
the medium. Atomic polarization (or spin \cite{LukinRMP03}) is described by an operator $\hat b(z,t)$ which removes one
atom's worth of internal excitation from the atomic ensemble localized in a thin-slice volume around a point in space-time in
the medium  \cite{RaymerJMO04}:
\begin{equation}\label{eq:bdef}
\hat b(z,t)=i\frac{\rho^{1/2}}{n}\sum_{\{\alpha\}_z} |1_\alpha\rangle\langle3_\alpha|
\,e^{i(\omega_p-\omega_s)t_\text{LAB}-i(k_p-k_s)z}
\end{equation}
where $\rho$ is linear atom density (atoms per mm), $n$ is the number of atoms in a slice, indexed by $\alpha$,
$|1\rangle$ and $|3\rangle$ denote the Raman-coupled atomic levels, $\omega_p$ and $\omega_s$ are the pump and Stokes
angular frequencies, while $k_p$ and $k_s$ are their wavevectors (see Fig.~\ref{fig:levels}). As long as the fraction
of excited atoms is small, $\hat b(z,t)$ is approximately a bosonic operator, $[\hat b(z,t),\hat b(z',t)^\dagger]\approx\delta(z-z')$.
The Stokes field is described by an envelope annihilation operator at each point in space along the pump beam $\hat a(z,t)$:
\begin{equation}\label{eq:adef}
\hat a(z,t)=\int d\omega \, e^{-i(\omega-\omega_s)t_\text{LAB}+ik_s z} \hat a(z,\omega).
\end{equation}
It is a bosonic operator, $[\hat a(z,t),\hat a(z,t')^\dagger]=\delta(t-t')$.
Finally, the pump field is given by its envelope $A_p(z,t)$ in the reference frame of the Stokes pulse, scaled to
unity, defined by
\begin{equation}\label{eq:Ap}
\frac{E_p(t-(z-L/2)\Delta\beta )}{E_{p,\text{max}}}=A_p(z,t)e^{-i\omega_p t+i k_p z} + \text{c.c.}
\end{equation}
where $E_p(t)$ is the pump electric field while $\Delta\beta$ is the difference of the inverse group velocities of the
pump and Stokes pulses (measured in picoseconds per milimeter)
\begin{equation}
\Delta\beta=\frac{1}{v_\text{gr,s}}-\frac{1}{v_\text{gr,p}}
\end{equation}
where  $v_\text{gr,p}$ is the pump group velocity. If the Stokes field is faster than the pump, then $\Delta\beta$ is
negative.

Using these definitions, the Raman scattering process is cast into the following pair of coupled equations
\cite{Raymer1DTheory81,RaymerJMO04}:
\begin{subequations}\label{eq:prop}
\begin{eqnarray}
\frac{\partial\hat a(z,t)}{\partial z}&=&g_0 A_p(z,t) \,\hat b^\dagger (z,t) \\
\frac{\partial\hat b(z,t)}{\partial t}&=&g_0 A_p(z,t) \,\hat a^\dagger(z,t) \label{eq:propb}
\end{eqnarray}
\end{subequations}
where $g_0$ is the coupling constant (measured in $(\text{mm}\text{ps})^{-1/2}$) defined as:
\begin{equation}\label{eq:g0def}
g_0=\rho^{1/2} \sqrt{\frac{\hbar\omega_s}{2\varepsilon_0 c}}\,E_{p,\text{max}}\kappa_1
\end{equation}
where $S$ is the pump beam cross-section while $\kappa_1$ is a coupling constant defined in \cite{Raymer1DTheory81}.
Since the pump and Stokes pulses are narrowband, higher order dispersion effects, possibly distorting the pulses, can be
neglected. On the other hand, in the atomic vapor one finds that the frequency difference between the interacting waves
is large enough for significant group velocity differences to be observed. A typical subnanosecond pump pulse can be
delayed by several times its duration with respect to the Stokes pulse, thus the presence of nonzero $\Delta\beta$
cannot be neglected \cite{Ji2005}.

\begin{figure}
  \center\includegraphics[width=0.45\textwidth]{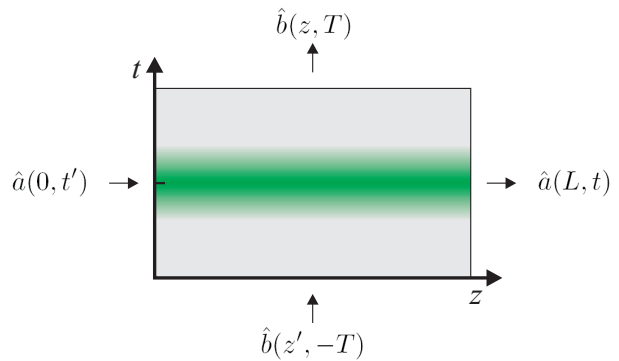}
  \caption{Pictorial representation of the Stokes scattering process in the Stokes reference frame, defined in \eeqref{eq:tStokes}.
  The medium entangles light field $\hat a$ with atoms $\hat b$. The local entanglement amplitude
  increases with a (fixed) local pump field strength $A_p(z,t)$, represented in the picture by a grey (green in color) stripe.
  Horizontal stripe corresponds to equal group velocities of pump and Stokes, $\Delta\beta=0$.
  }
  \label{fig:Stokes}
\end{figure}
We proceed with solving \eeqref{eq:prop} by integrating over a finite spatial-temporal region in which the interaction
takes place, thus formulating the problem as a quantum scattering problem. This approach, pictorially represented in
Fig.~\ref{fig:Stokes} requires introduction of a suitable scattering matrix, which in the case of Stokes scattering is
composed of the Green functions of \eeqref{eq:prop}. We choose to introduce them over a space-time region $0<z<L$ and
$-T<t<T$, in the following way:
\begin{subequations}\label{eq:io}
\begin{align}
\hat a(L,t)=&\int_{-T}^{T} dt' C_a(t,t') \hat a(0,t') \nonumber\\
&+ \int_{0}^{L} dz' S_a(t,z') \hat b^\dagger(z',-T) \\
\hat b(z,T)=&\int_{0}^{L} dz' C_b(z,z') \hat b(z',-T) \nonumber\\
&+ \int_{-T}^{T} dt' S_b(z,t') \hat a^\dagger(0,t')
\end{align}\end{subequations}
We can simplify the above input-output relations using an extended Bloch-Messiah reduction theorem \cite{BraunsteinBM},
which we derive in the Appendix~\ref{app:BM}. This theorem states that since the Green functions for this problem must
describe a unitary operator transformation, they possess common singular vectors and related singular values
$\cosh\zeta_n$ and $\sinh\zeta_n$:
\begin{subequations}\label{eq:CSdecomp}
\begin{align}
C_a(t,t')=&\sum_n \cosh\zeta_n \cdot \psi^{(out)*}_n(t)\, \psi^{(in)}_n(t') \label{eq:CSdecompCa}\\
C_b(z,z')=&\sum_n \cosh\zeta_n \cdot \phi^{(out)*}_n(z)\, \phi^{(in)}_n(z') \label{eq:CSdecompCb}\\
S_a(t,z')=&\sum_n \sinh\zeta_n \cdot \psi^{(out)*}_n(t)\, \phi^{(in)*}_n(z') \label{eq:CSdecompSa}\\
S_b(z,t')=&\sum_n \sinh\zeta_n \cdot \phi^{(out)*}_n(z)\, \psi^{(in)*}_n(t'), \label{eq:CSdecompSb}
\end{align}
\end{subequations}
where $\zeta_n$ are real numbers while $\psi^{(out)}_n(t)$, $\phi^{(out)}_n(z)$, $\psi^{(in)}_n(t)$ and
$\phi^{(in)}_n(z)$ are complex-valued functions which form separate orthonormal bases for the output and
input, light and atomic mode-function spaces respectively
\footnote{The singular value decomposition is, especially in case of decomposing wavefunction of bipartite system
$\Psi(z,z')$, better known as a Schmidt decomposition. For a review of this algebraic representation of an arbitrary
rectangular matrix, see \cite{SVDReview}.}:
\begin{equation}\label{eq:ortho}\begin{aligned}
\int_{-T}^{T} dt\,\psi^{(in)*}_n(t)  \cdot \psi^{(in)}_m(t)  &= \delta_{n,m} \\
\int_{0}^{L}  dz\,\phi^{(in)*}_n(z)  \cdot \phi^{(in)}_m(z)  &= \delta_{n,m} \\
\int_{-T}^{T} dt\,\psi^{(out)*}_n(t) \cdot \psi^{(out)}_m(t) &= \delta_{n,m} \\
\int_{0}^{L}  dz\,\phi^{(out)*}_n(z) \cdot \phi^{(out)}_m(z) &= \delta_{n,m}. \\
\end{aligned}\end{equation}
We insert the Green functions in the form given in \eeqref{eq:CSdecomp} into \eeqref{eq:io}. Then we express input and output field
annihilation operators $\hat a(0,t)$, $\hat b(z,-T)$ by mode expansions and define output operators $\hat a(L,t)$ and $\hat b(z,T)$
analogously
\begin{equation}\label{eq:abinoutexp}
\begin{aligned}
\hat a(0,t) &= \sum_n \hat a^{(in)}_n  \,\psi^{(in)*}_n(t)   \\
\hat a(L,t) &= \sum_n \hat a^{(out)}_n \,\psi^{(out)*}_n(t)  \\
\hat b(z,-T)&= \sum_n \hat b^{(in)}_n  \,\phi^{(in)*}_n(z)   \\
\hat b(z,T) &= \sum_n \hat b^{(out)}_n \,\phi^{(out)*}_n(z),
\end{aligned}
\end{equation}
where we have introduced operators $\hat a^{(in)}_n$, $\hat a^{(out)}_n$, $\hat b^{(in)}_n$ and $\hat b^{(out)}_n$. Thanks to
orthonormality of the respective mode functions expressed in \eeqref{eq:ortho}, these operators obey canonical commutational relations:
\begin{align}
[\hat a^{(out)}_n,\hat a^{(out)\dagger}_m]&=\delta_{nm} & [\hat b^{(out)}_n,\hat b^{(out)\dagger}_m]&=\delta_{nm}
\end{align}
and the operators with superscript $(in)$
obey analogous commutational relations. Therefore $\hat a^{(in)}_n$, $\hat a^{(out)}_n$, $\hat b^{(in)}_n$ and $\hat
b^{(out)}_n$ are bosonic operators, which annihilate excitations in respective light and atomic modes. Using
\eeqref{eq:CSdecomp}, \eqref{eq:ortho} and \eqref{eq:abinoutexp}, \eeqref{eq:io} can be cast into the following form:
\begin{subequations}\label{eq:mmsqueezing}
\begin{eqnarray}
\hat a^{(out)}_n&=&\hat a^{(in)}_n \cosh\zeta_n + \hat b^{(in)\dagger}_n \sinh\zeta_n \\
\hat b^{(out)}_n&=&\hat b^{(in)}_n \cosh\zeta_n + \hat a^{(in)\dagger}_n \sinh\zeta_n.
\end{eqnarray}
\end{subequations}
This expresses independent two-mode squeezing processes, and is the main result of this part of our study. Each of
these elementary processes squeezes the optical field in a particular temporal mode $\psi^{(in)}_n(t)$ with the atomic
polarization in a particular longitudinal spatial mode $\phi^{(in)}_n(z)$, producing fields in optical temporal modes
$\psi^{(out)}_n(t)$ that are pair-wise entangled with the internal states of atoms described by collective spatial modes
$\phi^{(out)}_n(z)$. The squeezing parameter equals $\zeta_n$. The noise in the sum and difference of quadratures
scales by a factor of $\exp(\zeta_n)$, corresponding to the mean number of excitations $\avg{\hat
n_n}=\avg{\hat a^{(out)\dagger}_n \hat a^{(out)}_n}=\avg{\hat b^{(out)\dagger}_n \hat b^{(out)}_n}=\sinh^2\zeta_n$.

Note that for a pump pulse of a given shape, characterized by duration $\tau_p$, for instance Gaussian:
\begin{equation}\label{eq:pump}
A_p(z,t)=\exp\left[-2\ln2\, \left(\frac{t-(z-L/2)\Delta\beta}{\tau_p}\right)^2\right],
\end{equation}
the time and space in equations \eqref{eq:prop} can be reduced to a dimensionless coordinates by measuring time in
units of pump pulse duration $\tau_p$ and space in units of distance over which temporal Stokes walk-off equals pump
duration $\tau_p/\Delta\beta$. This way the pump envelope $A_p(\Delta\beta z/\tau_p,t/\tau_p)$ becomes a fixed
function. Note that for $T\gg\tau_p$ the Green functions do not depend on $T$ since there is no longer any interaction
taking place for large $t$. Thus the Green functions defined in \eeqref{eq:io} can depend only on medium length,
coupling strength, and pulse shape. Both can be brought into a form of two dimensionless parameters:
\begin{subequations}\begin{align}
\Delta&=\frac{L\Delta\beta}{\tau_p} \\
\Gamma&=g_0 \sqrt{L\tau_p}\,.
\end{align}\end{subequations}
The first is the temporal walkoff between pump and Stokes field, measured in pump pulse durations, while the second
measures the squeezing strength and is proportional to the squeezing parameter in the case of zero dispersion
\cite{Raymer1DTheory81,RaymerJMO04}. Therefore there is a congruence between the solutions and mode functions for the
Stokes scattering problem, as long as the pump pulse shapes are congruent and the values of the parameters $\Gamma$ and
$\Delta$ are the same.

Finally, let us note that the characteristic output modes $\psi^{(out)}_n(t)$ and $\phi^{(out)}_n(t)$ we obtained with
the help of the Bloch-Messiah reduction are identical to previously obtained atomic and field coherence modes
\cite{Raymer3DTheory85,RaymerJMO04,RaymerTemporal89}. The latter were introduced as the eigenmodes of the first-order
coherence functions for atoms $\avg{\hat b(z,T)\hat b^\dagger(z',T)}$ and field $\avg{\hat a(L,t)\hat
a^\dagger(L,t')}$. Calculating these functions explicitly using equations \eqref{eq:CSdecomp} we obtain:
\begin{align}
\avg{\hat a(L,t)\hat a^\dagger(L,t')}=&
\int_{0}^{L} dz' S_a(t,z') S^*_a(t',z') \nonumber \\
=&\sum_n \avg{\hat n_n} \,\psi^{(out)*}_n(t)\, \psi^{(out)}_n(t')
\\
\avg{\hat b(z,T)\hat b^\dagger(z',T)}=&
\int_{-T}^{T} dt' S_b(z,t') S^*_b(z',t') \nonumber \\
=&\sum_n \avg{\hat n_n} \,\phi^{(out)*}_n(t)\, \phi^{(out)}_n(t')
\end{align}
as in \cite{Raymer3DTheory85,RaymerJMO04}. In \cite{RaymerJMO04} also the characteristic input modes were introduced as
those evolving into a particular pair of output modes, which is another way of obtaining the decomposition given in
\eeqref{eq:CSdecomp}.

\section{Numerical Results}\label{sec:StokesNumeric}
To illustrate the meaning of the above results and provide the decomposition into independent entanglement processes in
some specific and realistic cases, we have numerically found Green functions defined in \eeqref{eq:io}. Calculations
were carried out for a medium of length $L=75\,$mm for various values of group velocity difference $\Delta\beta$,
coupling strength $g_0$ and Gaussian pump pulse of FWHM duration $\tau_p=200\,$ps as defined in \eeqref{eq:pump}.

As the equations of motion \eqref{eq:prop} are linear, they are identical to the classical equations for evolution of
the Stokes field and the atomic polarization in the case of stimulated Raman scattering, in which annihilation
operators $\hat a(z,t)$ and $\hat b(z,t)$ are replaced by classical field $\alpha(z,t)$ and polarization amplitudes
$\beta(z,t)$.  Therefore we can compute matrix approximations to the Green functions $C_a(t,t')$, $C_b(z,z')$,
$S_a(t,z')$ and $S_b(z,t')$ using standard methods developed in nonlinear optics. To accomplish this, equations
\eqref{eq:prop} were solved numerically for a complete orthonormal set of initial conditions. First we put
$\alpha(0,t)=0$ for every $t$ except for a single point on a computational grid $t=t'$ where $\alpha(0,t')=1$, and
$\beta(z,-T)=0$ for every $z$. Performing finite integration steps over the space-time grid of 200$\times$200 points
where the interaction occurs, we computed $\alpha(L,t)$ and $\beta(z,T)$. For the initial conditions chosen, they are
equal to $C_a(t,t')$ and $S_b(z,t')$, respectively. Repeating this calculation for each $t'$, we constructed a matrix
approximation of these Green functions. Then analogously starting from $\alpha(0,t)=0$, and $\beta(z,-T)=0$ except for
a single point on a computational grid where $\beta(z',-T)=1$, we have constructed $C_b(z,z')$, $S_a(t,z')$. Next we
numerically performed the singular value decomposition \cite{SVDReview} of these matrices to obtain the singular values $\sinh \zeta_n$ and $\cosh
\zeta_n$ as well as associated vectors $\psi^{(in)}_n(t)$, $\phi^{(in)}_n(t)$, $\psi^{(out)}_n(t)$ and
$\phi^{(out)}_n(t)$ as defined in \eeqref{eq:CSdecomp}. We have verified that our simulation is accurate by repeating
it on a grid two times finer in each dimension. The differences between the results were of the order of precision with
which singular value decomposition can be computed. We have also verified that our numerical calculation reconstructs
the analytical results for square pump pulse without dispersion \cite{RaymerJMO04}.

\begin{figure}\center
  \includegraphics[width=0.45\textwidth]{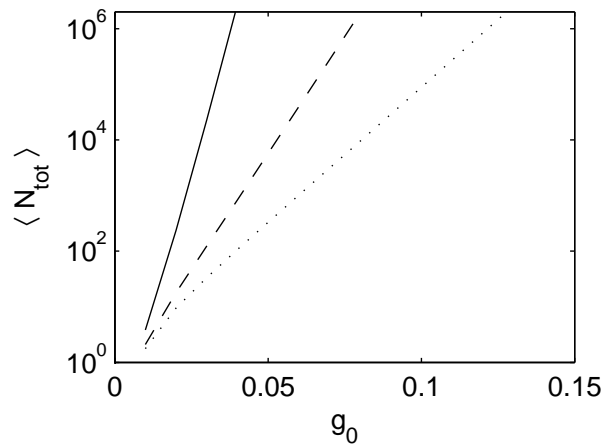}
  \caption{Mean number of Stokes photons as a function of coupling strength $g_0$ [(mm ps)$^{-1/2}$].
  Three lines are the results of computations for different $\Delta\beta$ equal $0$ (solid line), $-10\,$ps/mm (dashed line)
  and $-30\,$ps/mm (dotted line).}
  \label{Ntotvsg}
\end{figure}
Numerical calculations confirm that the total number of photons
\begin{equation}
\avg{N_\text{tot}}=\sum_n\avg{\hat n_n}=\sum_n \sinh^2\zeta_n .
\end{equation}
is, to a good approximation, an exponential function of coupling $g_0$, within the range plotted in Fig.~\ref{Ntotvsg}. However, we
find a strong dependence of $\avg{N_\text{tot}}$ on the group velocity difference between pump and Stokes pulses
$\Delta\beta$. With increasing $\Delta\beta$ the total photon number $\avg{N_\text{tot}}$ quickly decreases. Using
results displayed in Fig.~\ref{Ntotvsg} we have found, for each $\Delta\beta$, the value of coupling $g_0$ for which
the total number of photons equals $\avg{N_\text{tot}}=10^6$. These values of coupling $g_0$ will be used in the
following examples.

\begin{figure}
  \center{\raisebox{3.5cm}{$\avg{\hat n_n}$}\hskip0mm\includegraphics[width=0.4\textwidth]{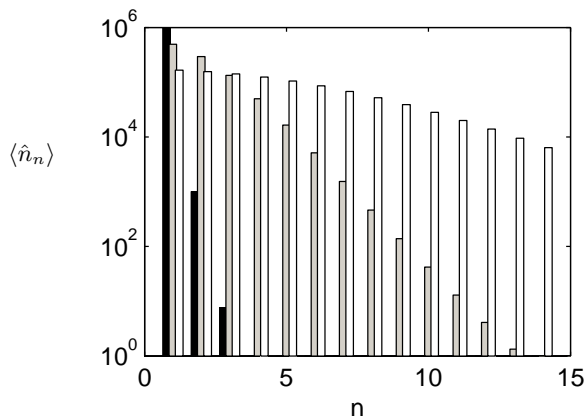}}
  \caption{Mean number of photons in each of characteristic output modes $\avg{\hat n_n}$ for 15 most-occupied modes.
  The histograms has been made for coupling yielding average total number of Stokes photons $\langle N_\text{tot}
  \rangle=10^6$ for different $\Delta\beta$ equal $0$ (black bar), $-10\,$ps/mm (grey bars)
  and $-30\,$ps/mm (empty bars).}
  \label{Nvsg}
\end{figure}
In Fig.~\ref{Nvsg} we plot the mean number of photons in each mode $\avg{\hat n_n}=\sinh^2\zeta_n$. In the case of no
dispersion $\Delta\beta=0$, the occupation of the first mode dominates by a factor of $10^3$ over the next, as known
previously \cite{RaymerStatistics82,WalmsleyStatistics83}. Strong domination, however, quickly comes to an end with increasing dispersion.
For $\Delta\beta=-30\,$ps/mm more than a dozen modes have occupancy equal within an order of magnitude.

\begin{figure}
  \center\includegraphics[width=0.5\textwidth]{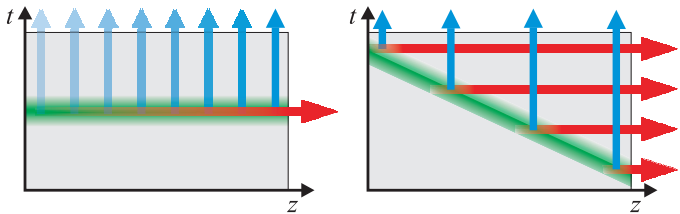} \caption{Pictorial representation of the Stokes scattering for
  zero group velocity difference $\Delta\beta$ (left) and large negative $\Delta\beta\ll-\tau_p/L$ (right). The grey
  box represents the spatio-temporal region of interaction, the dark grey (green in color) stripe is the pumped region, the
  horizontal (red) arrows are the Stokes emission, while the vertical (blue) arrows represent atomic polarization. The picture has
  been drawn in the Stokes reference frame.} \label{fig:Stokes2}
\end{figure}
\begin{figure*}
\begin{tabular}{ccc}
    $\Delta\beta=0$ &     $\Delta\beta=-10\,$ps/mm &     $\Delta\beta=-30\,$ps/mm \\
  \includegraphics[scale=0.75]{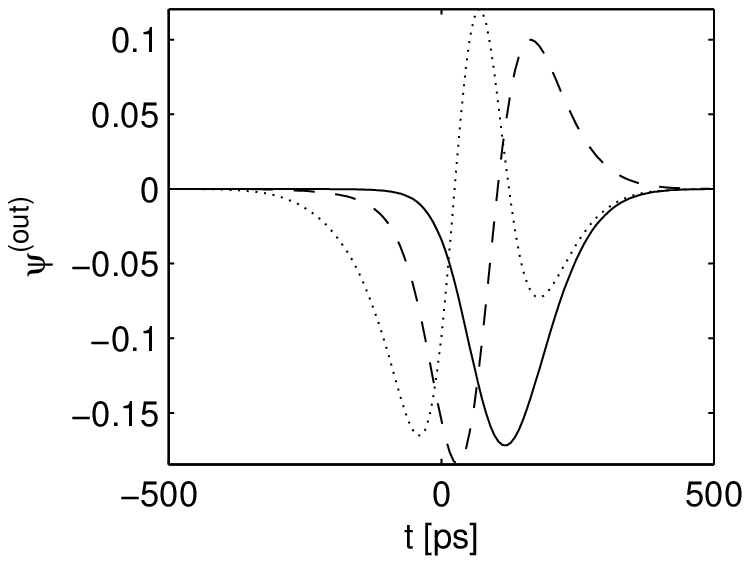} & \includegraphics[scale=0.75]{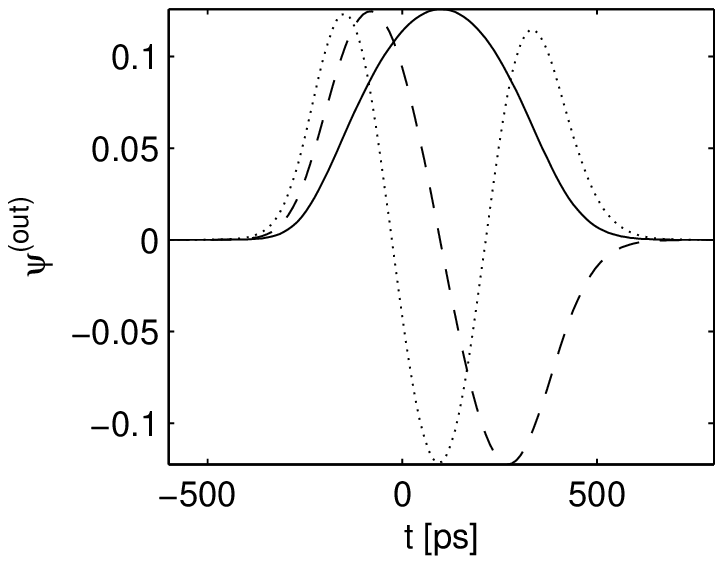} & \includegraphics[scale=0.75]{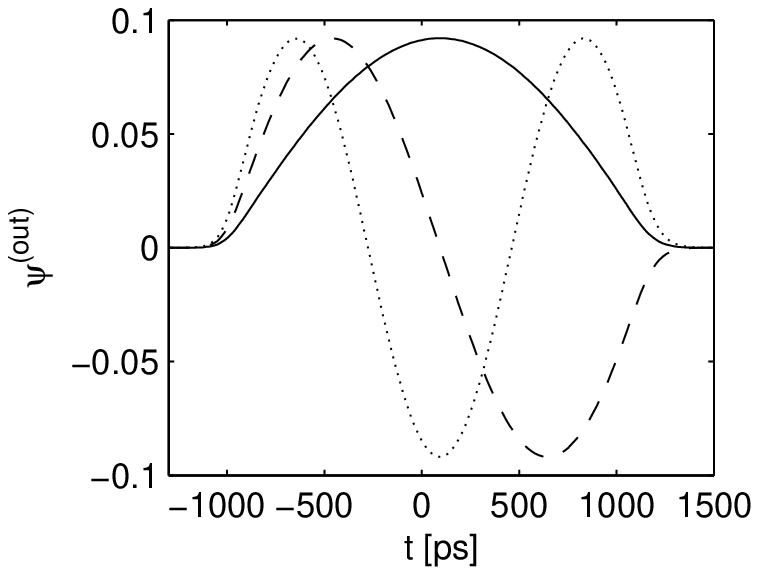} \\
  \includegraphics[scale=0.75]{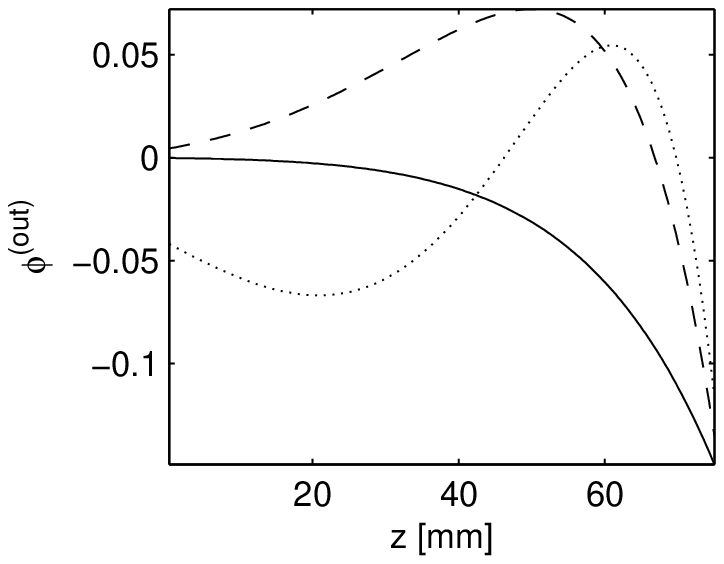} & \includegraphics[scale=0.75]{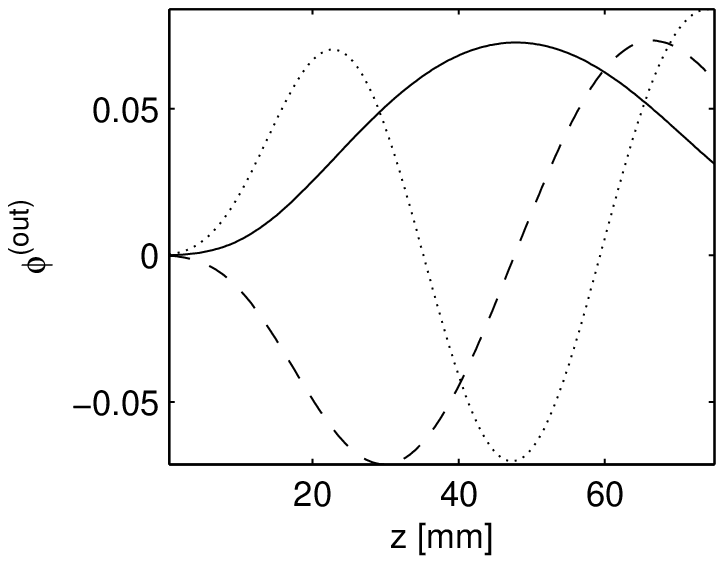} & \includegraphics[scale=0.75]{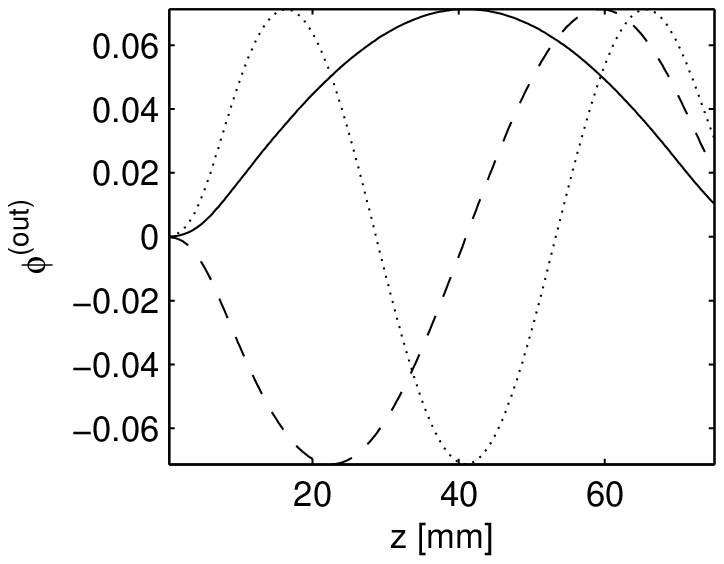}
\end{tabular}
  \caption{Optical $\psi^{(out)}_n(t)$ and atomic $\phi^{(out)}_n(t)$ output mode functions for three most excited modes,
  $n=1$  (solid line),  $n=2$ (dashed line) and $n=3$ (dotted line) for various group velocity differences $\Delta\beta$.}
  \label{psiout}
\end{figure*}
The physical reason behind this result can be traced in Fig.~\ref{fig:Stokes2}, where we depict a space-time region
where the interaction occurs. We mark the pumped area as the dark grey (green) region. In the case of zero group velocity
difference, $\Delta\beta=0$, the first few Stokes photons propagate through the pumped region, stimulating subsequent
scattering acts and inducing coherence between atomic polarization in various regions along the cell. The process is
almost single mode \cite{RaymerStatistics82,RaymerJMO04}. On the other hand, in the case of high group velocity
difference, the spatial-temporal interaction region can be divided into several regions, the number of such regions
being roughly the ratio of the total temporal walkoff to the pump duration $\Delta=L\Delta\beta/\tau_p$. These regions are
weakly coupled, since the photons scattered in the distinct parts of the cell do not have a chance to meet each other.
Thus several uncorrelated temporal modes take part in the process and become significantly occupied.

This latter phenomenon complicates the possible applications. For nonzero dispersion one must afford means, such as for
example balanced homodyne detection, for separating distinct output modes in order to extract a pure squeezed state of
light and matter. Whether in the single or multimode squeezing regime, an efficient detection of Stokes light is
necessary for quantum communication and computation. This is possible by means of balanced homodyne detection using a
local oscillator in the shape of the characteristic Stokes mode.  In Fig.~\ref{psiout} we plot the temporal shape of
the most occupied output field modes for the Gaussian pump centered at $t=0$ defined in \eqref{eq:pump}. The dominant
mode has a single peak, which is time delayed relative to pump peak \cite{RaymerTemporal89}. The modes resemble in shape the
eigenmodes of a harmonic oscillator. They are also strongly affected by the group velocity difference. With increasing
$\Delta\beta$, the duration of the fundamental mode increases while the time delay relative to the pump pulse in the
middle of the medium decreases. Also for large $\Delta\beta$ all the characteristic modes have significant intensity
over the same period of time, spanning approximately one pump duration plus temporal walkoff between pump and the
Stokes $\tau_p+L\Delta\beta=\tau_p(1+\Delta)$.

Note that in the limit of zero dispersion, $\Delta\beta=0$, the Green functions calculated numerically can be found in
a closed form using Laplace-transform techniques \cite{Raymer1DTheory81}, even for the full three-dimensional case
\cite{Raymer3DTheory85}. The results presented here for zero dispersion are recovered by applying numerical
diagonalization directly to those functions \cite{RaymerJMO04}. Finally let us note that the output characteristic modes
can be used to reconstruct the statistics of fluctuating Stokes pulse shapes \cite{RaymerTemporal89}.

\section{Photon count statistics}\label{sec:PC}
Using the results of numerical calculations, we were able to compute the photon-count statistics for the Stokes pulse.
Let us recall that one mode of the pair comprising a two-mode squeezed state exhibits thermal statistics. This is the
case of any particular output field mode $\psi^{(out)}_n(t)$ by itself. Also, each of the output field modes is
independent, since they evolve separately from vacuum. Thus the Stokes pulse has multimode thermal statistics
\cite{RaymerStatistics82}:
\begin{equation}
p(n)=\frac{1}{2\pi} \int dk  \frac{e^{-ikn}}{\prod_n (1-ik\avg{\hat n_n})}
\end{equation}

\begin{figure}
  \center\includegraphics[scale=0.8]{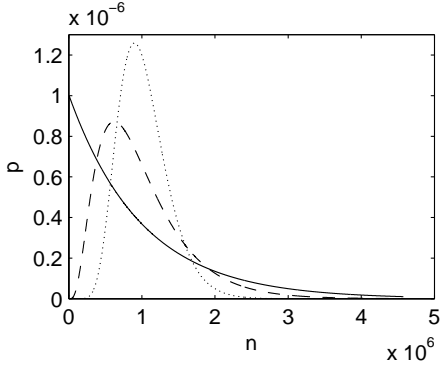}
  \caption{Photon count probability distribution $p(n)$ of Stokes light pulses.
  Three lines are the results of computations for different $\Delta\beta$ equal $0$ (solid line), $-10\,$ps/mm (dashed line)
  and $-30\,$ps/mm (dotted line).}
  \label{p_n}
\end{figure}
\begin{figure}
  \center\includegraphics[scale=0.8]{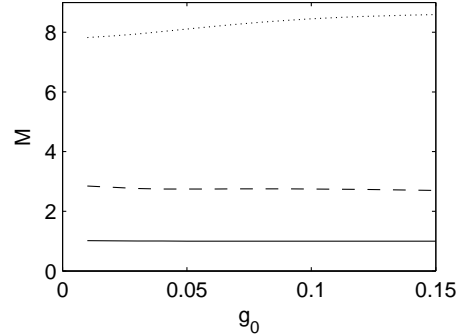}
  \caption{Equivalent number of independent thermal modes $\mathcal{M}$ as a function of coupling strength $g_0$ [(mm ps)$^{-1/2}$].
  Three lines are the results of computations for different $\Delta\beta$ equal $0$ (solid line), $-10\,$ps/mm (dashed line)
  and $-30\,$ps/mm (dotted line).}
  \label{Mvsg}
\end{figure}
We plot $p(n)$ in Fig.~\ref{p_n} for total photon number $\avg{N_\text{tot}}=10^6$ and various values of group velocity
difference $\Delta\beta$.

Following the analogy with thermal light, we have also computed and plot in Fig.~\ref{Mvsg} the equivalent number of
modes $\mathcal{M}$. It is defined as a number of independent thermal modes of equal intensity that measured together
reproduce the Stokes fluctuation-to-mean ratio:
\begin{equation}
\mathcal{M}=\frac{\avg{N_{tot}}^2}{\sum_n \avg{\hat n_n}^2 .}
\end{equation}
As can be seen in Fig.~\ref{Mvsg}, the equivalent mode number $\mathcal{M}$ depends mostly on $\Delta\beta$, which is
almost independent of coupling strength $g_0$. This is consistent with the intuitive way of understanding Stokes
scattering we propose in Fig.~\ref{fig:Stokes2}.

\section{Atomic polarization readout}\label{sec:aStokes}
As shown in Section~\ref{sec:StokesTheory}, the optical Stokes pulse contains only one part of the squeezed state.
Alone it has thermal statistics and possesses no distinct quantum features. Entanglement can  be observed only if the
field counterpart --- the atomic polarization --- is acted upon. A convenient to way process the atomic $\hat b(z)$ field is
to translate its statistics into the optical field. This is done in the anti-Stokes scattering process, during which a
portion of atomic polarization gives rise to an anti-Stokes pulse. The equations governing this process are similar to
Eqs.~\eqref{eq:prop} in the anti-Stokes reference frame \cite{RaymerJMO04}:
\begin{subequations}\label{eq:aprop}
\begin{eqnarray}
\frac{\partial\hat c(z,t)}{\partial z}&=&g'_0 A'_p(z,t)\hat d(z,t) \\
\frac{\partial\hat d(z,t)}{\partial t}&=&-g'^*_0 A'^*_p(z,t)\hat c(z,t) \label{eq:apropb}
\end{eqnarray}
\end{subequations}
where $\hat c(z,t)$ is the anti-Stokes field annihilation operator, defined by an equation analogous to
\eqref{eq:adef}, $\hat d(z,t)$ is the atomic polarization operator, defined by an equation analogous to
\eqref{eq:bdef}, $g'_0$ is the readout coupling defined analogously to \eqref{eq:g0def}, while $A'_p(z,t)$ is the
readout pump field envelope. The definition of the $A'_p(z,t)$ is analogous to that of the Stokes pump field envelope
$A_p(z,t)$ given in \eeqref{eq:Ap}, but with a different peak amplitude and a different group velocity difference,
which we will designate by $\Delta\beta'$:
\begin{equation}
\Delta\beta'=\frac{1}{v_\text{gr,a}}-\frac{1}{v_\text{gr,p}}
\end{equation}
where $v_\text{gr,a}$ is the group velocity of the anti-Stokes wave, while $v_\text{gr,p}$ is the group velocity of the
readout pump.

Again, we formulate the readout problem as a quantum scattering problem, analogously to the situation illustrated in
Fig.~\ref{fig:Stokes}. This time the scattering matrix is composed of four Green functions for the anti-Stokes
scattering equations \eqref{eq:aprop}. We define them analogously to those defined in \eeqref{eq:io}  for the Stokes
scattering:
\begin{subequations}\label{eq:aio}
\begin{eqnarray}
\hat c(L,t)&=&\int_{-T}^T dt' C_c(t,t') \hat c(0,t') \nonumber\\
&&+ \int_0^L dz' S_c(t,z') \hat d(z',-T) \\
\hat d(z,T)&=&\int_0^L dz' C_d(z,z') \hat d(z',-T) \nonumber\\
&&- \int_{-T}^T dt' S_d(z,t') \hat c(0,t')
\end{eqnarray}\end{subequations}
Here $\hat c(0,t)$ is the optical field operator (usually representing the vacuum) entering the vapor during readout,
and  $\hat d(z,-T)$ is the atomic ensemble operator just prior to the readout.

We can cast these relations into multiple, independent beamsplitter transformations (see Appendix \ref{app:BS}). Since
the Green functions for this problem must describe a unitary transformation, they possess common singular vectors and
related singular values, which we parameterize by $\eta_n$:
\begin{subequations}\label{eq:BCSdecomp}
\begin{align}
C_c(t,t')=&\sum_n \sqrt{1-\eta_n} \cdot \Psi^{(out)*}_n(t)\, \Psi^{(in)}_n(t') \\
C_d(z,z')=&\sum_n \sqrt{1-\eta_n} \cdot \Phi^{(out)*}_n(z)\, \Phi^{(in)}_n(z') \\
S_c(t,z')=&\sum_n \sqrt{\eta_n}   \cdot \Psi^{(out)*}_n(t)\, \Phi^{(in)}_n(z') \\
S_d(z,t')=&\sum_n \sqrt{\eta_n}   \cdot \Phi^{(out)*}_n(z)\, \Psi^{(in)}_n(t').
\end{align}
\end{subequations}
The readout modes $\Psi^{(out)}_n(t)$, $\Phi^{(out)}_n(z)$, $\Psi^{(in)}_n(t)$ and $\Phi^{(in)}_n(z)$ form orthonormal
bases in the intput/output, light and atomic modefunction spaces:
\begin{equation}\label{eq:aortho}\begin{aligned}
\int_{-T}^{T} dt\,\Psi^{(in)*}_n(t)  \cdot \Psi^{(in)}_m(t)  &= \delta_{n,m} \\
\int_{0}^{L}  dz\,\Phi^{(in)*}_n(z)  \cdot \Phi^{(in)}_m(z)  &= \delta_{n,m} \\
\int_{-T}^{T} dt\,\Psi^{(out)*}_n(t) \cdot \Psi^{(out)}_m(t) &= \delta_{n,m} \\
\int_{0}^{L}  dz\,\Phi^{(out)*}_n(z) \cdot \Phi^{(out)}_m(z) &= \delta_{n,m}. \\
\end{aligned}\end{equation}
We use these functions as mode bases in respective spaces and decompose the annihilation operators
$\hat c(0,t)$, $\hat c(L,t)$, $\hat d(z,-T)$ and $\hat d(z,T)$:
\begin{subequations}\label{eq:cdinoutexp}
\begin{align}
\hat c(0,t) &= \sum_n \hat c^{(in)}_n  \,\Psi^{(in)*}_n(t)   \\
\hat c(L,t) &= \sum_n \hat c^{(out)}_n \,\Psi^{(out)*}_n(t)  \\
\hat d(z,-T)&= \sum_n \hat d^{(in)}_n  \,\Phi^{(in)*}_n(z)   \\
\hat d(z,T) &= \sum_n \hat d^{(out)}_n \,\Phi^{(out)*}_n(z) .
\end{align}
\end{subequations}
Where we have introduced annihilation operators of the input and output characteristic modes $\hat c^{(in)}_n$, $\hat
d^{(in)}_n$, $\hat c^{(out)}_n$ and $\hat d^{(out)}_n$. Again, it turns out that the output operators $\hat
c^{(out)}_n$ and $\hat d^{(out)}_n$ are independent bosonic operators. Same is true for input operators $\hat
c^{(in)}_n$ and $\hat d^{(in)}_n$.

These operators annihilate excitations of modes that undergo transformations analogous to a beamsplitter
transformation:
\begin{subequations}\label{eq:areducedio}
\begin{eqnarray}
\hat c^{(out)}_n&=&\sqrt{1-\eta_n}\,\hat c^{(in)}_n + \sqrt{\eta_n}\,\hat d^{(in)}_n \\
\hat d^{(out)}_n&=&\sqrt{1-\eta_n}\,\hat d^{(in)}_n - \sqrt{\eta_n}\,\hat c^{(in)}_n .
\end{eqnarray}
\end{subequations}
The real parameters $\eta_n$ have the meaning of the readout efficiency. Modes with $\eta_n=1$ are completely translated
from the atomic field into the optical field while spatial modes with $\eta_n=0$ remain unaffected by the anti-Stokes
scattering process.

\begin{figure}
  \center\includegraphics[width=0.45\textwidth]{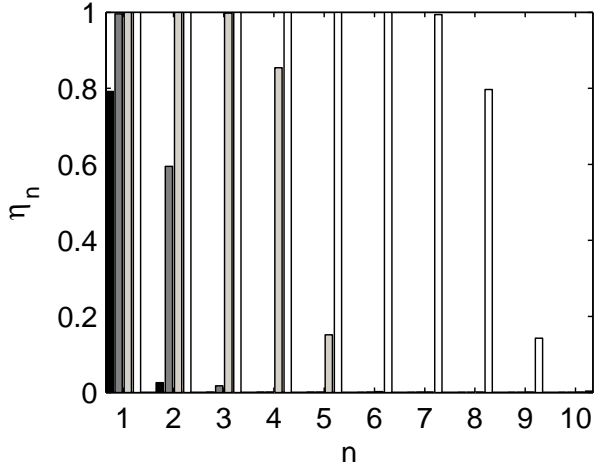}
  \caption{Readout efficiency $\eta_n$ for the most efficiently read modes in anti-Stokes scattering process for $\Delta\beta'=0$
  and coupling $g'_0$ equal $0.01,\,0.02,\,0.05,\,0.1\,(\text{ps}\,\text{mm})^{-1/2}$, for black, grey and empty bars respectively.}
  \label{Rvsn}
\end{figure}
We have numerically found forms for the Green functions defined in \eeqref{eq:aio} for similar conditions used to
calculate the Raman process in earlier sections. Results are first presented for the case of a medium of length
$L=75\,$mm with zero group velocity difference $\Delta\beta'$, various coupling strengths $g'_0$ and Gaussian pump
pulse $A'_p(z,t)$ of FWHM duration $\tau'_p=200\,$ps given in \eeqref{eq:pump}. In Fig.~\ref{Rvsn} we plot the
calculated $\eta_n$. It can be seen that above a certain value of coupling $g'_0$, many modes have unit readout
efficiency $\eta_n=1$, i.e. are completely translated into the optical field. Those modes are degenerate, thus within a
subspace spanned by their mode functions we can change the basis, i.e. the modal functions, without affecting relations
\eqref{eq:areducedio}.

Let us focus on the case that the readout takes place just following the Raman scattering process described in earlier
sections; in this case  $\hat d(z,-T)$ is given by the output operator $\hat b(z,T)$ in \eeqref{eq:io}:
\begin{equation}
\hat d(z,-T)=e^{i\Delta k\, z}\,\hat b(z,T)
\end{equation}
where $\Delta k$ is the wavevector mismatch in the time-delayed four-wave mixing process:
\begin{equation}
\Delta k = k_p-k_s+k_a-k_{p'}\,,
\end{equation}
where subscript $p$, $p'$, $s$, $a$ denote Stokes pump and anti-Stokes pump, Stokes and anti-Stokes signal waves
respectively. For efficient readout the phase-matching condition $\Delta k=0$ must be satisfied. Below we will assume
perfect phasematching. It is possible to satisfy this by proper choice of pump frequencies.

 Using the expansions \eqref{eq:abinoutexp} and \eqref{eq:cdinoutexp} we find that the characteristic anti-Stokes input
operators $\hat d^{(in)}_n$ are related to the characteristic Stokes output operators $\hat b^{(out)}_n$ by an unitary
transformation expressing the change of basis from $\phi^{(out)}_n(z)$ to $\Phi^{(in)}_n(z)$:
\begin{equation}
\hat d^{(in)}_m=\sum_n \mathbf{U}_{mn} \hat b^{(out)}_n
\end{equation}
where the transformation coefficients are:
\begin{equation}
\mathbf{U}_{mn}=\int_0^L dz\, \phi^{(out)*}_n(z)e^{-i\Delta k z}\cdot\Phi^{(in)}_m(z).
\end{equation}
Let us consider the readout of the first characteristic atomic mode of the Stokes scattering $\phi^{(out)}_1(z)$ whose
quantum statistics, before the readout, is contained in the operator $\hat b^{(out)}_1$. It can be decomposed in the
characteristic anti-Stokes input mode base $\Phi^{(in)}_n(z)$, according to the above formulas. The readout has unit
efficiency if the atomic mode $\phi^{(out)}_1(z)$ decomposes only into readout characteristic input modes
$\Phi^{(in)}_n(z)$ for which $\eta_m=1$. In other words, if for all $m$ for which $\mathbf{U}_{m1}\ne 0$ the readout
efficiencies $\eta_m$ equal one, then the readout has unit efficiency.

There is however a more straightforward way to analyze a readout of an excitation in a particular mode. Assuming
perfect phase matching, $\Delta k=0$, so $\hat d(z,-T)=\hat b(z,T)$ we substitute input atomic operator $\hat d(z,-T)$
in \eeqref{eq:aio} by $\hat b(z,T)$ in an expanded form given by \eeqref{eq:abinoutexp}:
\begin{subequations}\label{eq:aioexp}
\begin{align}
\hat c(L,t)&=\sum_n \sigma^*_n(t) \hat b^{(out)}_n + \int_{-T}^T dt' C_c(t,t') \hat c(0,t')\\
\hat d(z,T)&=\sum_n \varepsilon^*_n(z) \hat b^{(out)}_n - \int_{-T}^T dt' S_d(z,t') \hat c(0,t')
\label{eq:aioexp:d}
\end{align}\end{subequations}
where we have introduced mode functions
\begin{subequations}\label{eq:defsigmaepsilon}\begin{align}
\sigma^*_n(t)     =&\int_0^L dz'\, S_c(t,z')\, \phi^{(out)*}_n(z') \\
\varepsilon^*_n(z)=&\int_0^L dz'\, C_d(z,z')\, \phi^{(out)*}_n(z').
\end{align}\end{subequations}
The equations \eqref{eq:aioexp} give yet another modal decomposition of the output atomic and light fields.
However this time the modal functions: $\sigma^*_n(t)$, $C_c(t,t')$, $\varepsilon^*_n(z)$ and $S_d(z,t')$
form othonormal bases in the joint atom-light space:
\begin{subequations}\label{eq:asingleortho}\begin{align}
\int_{-T}^{T} dt\, \sigma^*_n(t)\sigma_m(t) + \int_0^L dz\,\varepsilon^*_n(z)\varepsilon_m(z)=\delta_{mn} \\
\int_{-T}^{T} dt\, \sigma^*_n(t) C_c(t,t')  - \int_0^L dz\,\varepsilon^*_n(z) S_d(z,t')=0.
\end{align}\end{subequations}
This is a consequence of unitarity of the relations \eqref{eq:aio} (see also Appendix~\ref{app:BS}).

Usually we want statistics of some particular atomic mode, say $\phi^{(out)}_1(z)$, to be completely transferred into
an optical field mode. This requirement is equivalent to requesting $\varepsilon_1(z)=0$ for every $z$, so that $\hat
d(z,T)$ in \eeqref{eq:aioexp:d} contains only zero-point, vaccum noise represented by $\hat c(0,t')$. The output light
mode $\sigma_1(t)$ can be computed in advance and detection apparatus tuned to it. Note, that if the transfer is
complete the $z$ integrals containing $\varepsilon_1(z)$ vanish in \eeqref{eq:asingleortho}. It follows that all other
output field modes must be orthogonal to $\sigma_1(t)$. In other words no crosstalk occurs, i.e. the polarization in
any other atomic mode does not give rise to any photons in the optical field mode $\sigma_1(t)$ we are presumably tuned
to. This does not prevent emission into other optical modes, however.

Therefore the readout problem can be reduced to a question of how to prepare the readout pump pulse in such a way that the
atomic polarization of interest, say in mode $\phi^{(out)}_1(z)$, is completely transferred into the optical field.

\begin{figure}
  \center\includegraphics[width=0.45\textwidth]{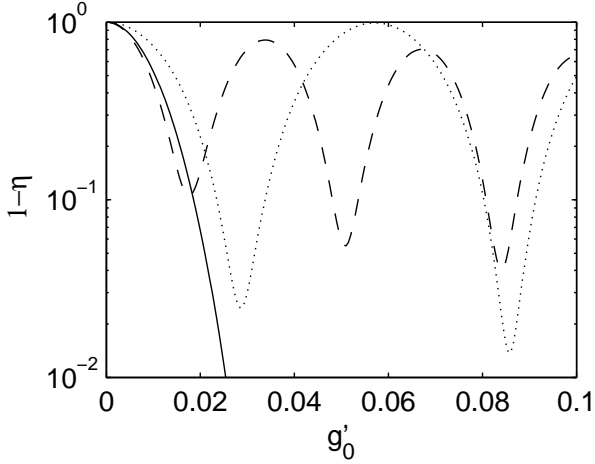}
  \caption{Fraction of residual atomic polarization $1-\eta$ not transferred to an optical field after the anti-Stokes scattering
  process as a function of the readout coupling $g_0\,[(\text{ps}\,\text{mm})^{-1/2}]$. Results for three group velocity
  differences $\Delta\beta'$ equal $0$ (solid line), $-10\,$ps/mm (dashed line) and $-30\,$ps/mm (dotted line) are displayed.}
  \label{Tvsg}
\end{figure}
\begin{figure*}\center
\begin{tabular}{cccc}
   & $\Delta\beta'=0$ &     $\Delta\beta'=-10\,$ps/mm &     $\Delta\beta'=-30\,$ps/mm \\
  \raisebox{2.3cm}{\small $\sigma_1(t)$}\hskip-4mm &
  \includegraphics[scale=0.7]{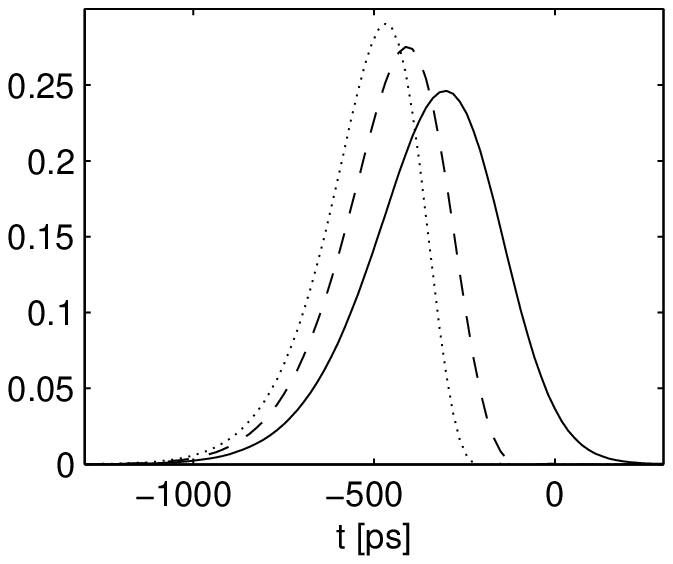} &
  \includegraphics[scale=0.7]{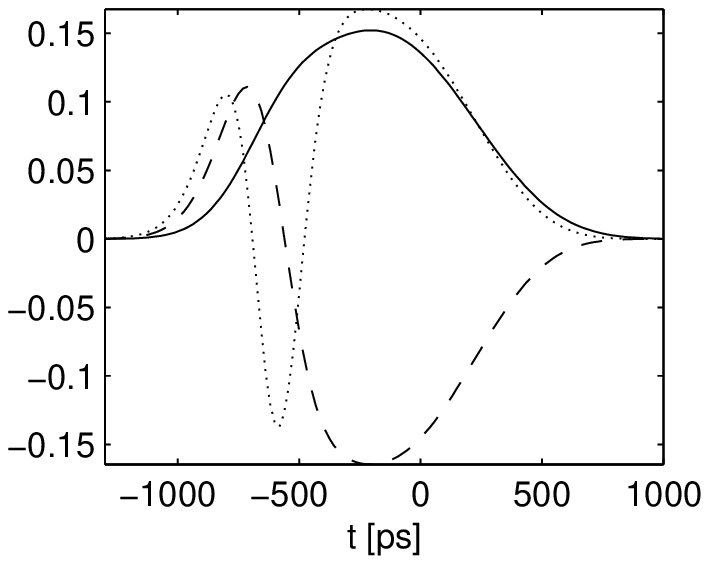} &
  \includegraphics[scale=0.7]{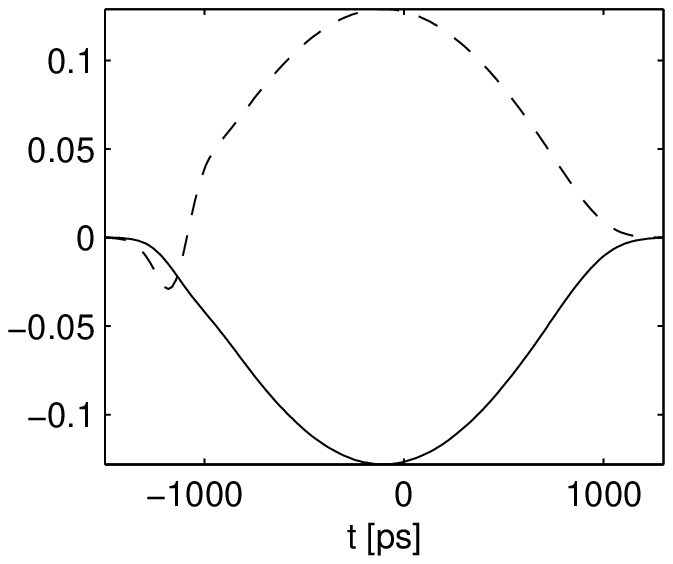} \\
   & $g'_0=0.04,0.07,0.1$ & $g'_0=0.018,0.05,0.084$ & $g'_0=0.028,0.086$ (mm ps)$^{-1/2}$
\end{tabular}
  \caption{Readout anti-Stokes light modes $\sigma_1(t)$ for fundamental atomic Stokes modes $\phi^{(out)}_1(z)$ for
  various group velocity difference $\Delta\beta=\Delta\beta'$ and coupling strengths corresponding to readout efficiency
  peaks for nonzero $\Delta\beta'$ (compare Fig.~\ref{Tvsg}). Solid, dashed and dotted lines correspond respectively to subsequent
  values of $g'_0$, listed below each panel.}
  \label{ROc}
\end{figure*}
We investigated this by simulating readout of an atomic mode $\phi^{(out)}_1(z)$ in an anti-Stokes interaction.
Calculations were carried out for a medium of length $L=75\,$mm for various values of group velocity differences,
assuming $\Delta\beta=\Delta\beta'$. For each particular value of this parameter, we have first computed the mode
function of the fundamental atomic output mode $\phi^{(out)}_1(z)$ in a way described in Sec.~\ref{sec:StokesNumeric},
with the Stokes coupling such that the total mean number of excitations $\avg{N_\text{tot}}=10^6$. Next we have
simulated the anti-Stokes scattering of this particular atomic mode for various anti-Stokes coupling strengths $g'_0$
and Gaussian pump $A'_p(z,t)$ of FWHM duration $\tau'_p=200\,$ps given in \eeqref{eq:pump}. We have quantified the
quality of the readout by computing the residual fraction of atomic polarization after the readout
\begin{equation}
1-\eta=\int_0^L dz |\varepsilon_1(z)|^2
\end{equation}
The results are shown in Fig.~\ref{Tvsg}. One can see that for zero group velocity difference $1-\eta$ rapidly falls
with increasing coupling $g'_0$ as predicted in \cite{RaymerJMO04}. The readout is virtually perfect for
$g'_0>0.04\,(\text{ps}\,\text{mm})^{-(1/2)}$. However for nonzero $\Delta\beta'$ the residual atomic polarization is of
the order of at least a percent. The readout inefficiency $1-\eta$ becomes a periodic function of coupling $g'_0$.  The
period increases with increasing $\Delta\beta'$, and the particular optimal values of $g'_0$ depend on the parameters
of the readout process as well as the initial atomic excitation mode function $\phi^{(out)}_1(z)$.

In Fig.~\ref{ROc} we plot the output field modes $\sigma_1(t)$ obtained in the readout simulations. In case of homodyne
detection of the output of the anti-Stokes readout process, those would be the optimal local oscillator shapes. For zero
group velocity difference $\Delta\beta'=0$ we observe that the output field mode becomes shorter with increasing
$g'_0$. For nonzero $\Delta\beta'$ the mode shapes $\sigma_1(t)$ corresponding to the subsequent minima of the function
$1-\eta(g'_0)$ depicted in the Fig.~\ref{Tvsg} show increasing number of nodes.

\begin{figure}[b]
  \center\includegraphics[width=0.5\textwidth]{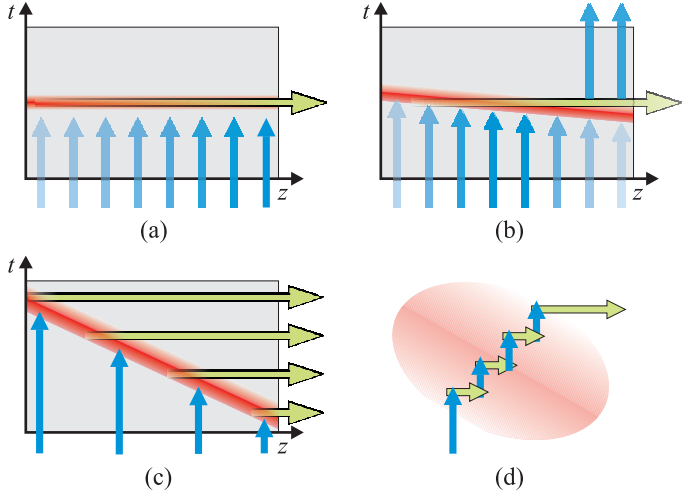}
  \caption{
   Pictorial representation of the anti-Stokes scattering for zero group velocity difference $\Delta\beta'$ (a), medium
  value of $\Delta\beta'$ (b) and high negative $\Delta\beta'\ll-\tau_p/L$ (c). The grey box represents the spatio-temporal region of
  interaction, the dark grey (red in color) stripe is the pumped region, the horizontal (green) arrows are the
  anti-Stokes emission, while the vertical (blue) arrows represent atomic polarization. The picture is drawn in
  the anti-Stokes reference frame. Panel (d) gives insight into oscillatory exchange of excitation between light and
  atoms.} \label{fig:antiStokes}
\end{figure}
The physics behind the readout process can be intuitively understood if one notes that integrating the propagation
equations \eqref{eq:aprop} over a spatiotemporal region $2\Delta z\times 2\Delta t$ where the fields are almost constant
gives a simple beamsplitter-like relation between input and output fields:
\begin{align}
\hat c(z+\Delta z,t)=& \hat c(z-\Delta z,t)\sqrt{1-|\gamma|^2}+\gamma\hat d(z,t-\Delta t)\frac{\sqrt{\Delta z}}{\sqrt{\Delta t}} \nonumber\\
\hat d(z,t+\Delta t)=& \hat d(z,t-\Delta t)\sqrt{1-|\gamma|^2}-\gamma^*\hat c(z-\Delta z,t) \frac{\sqrt{\Delta t}}{\sqrt{\Delta z}}
\end{align}
where $\gamma=2g'_0 A'_p(z,t) \sqrt{\Delta z \Delta t}$. Therefore we can visualize the scattering process as a square
network of virtual beamsplitters, each having different reflectivity $\sqrt\gamma$. In Fig.~\ref{fig:antiStokes} we
represent the pumping intensity $A'_p(z,t)$ and the beams representing atomic polarization and optical field, which
propagate in above mentioned network.

For zero group velocity difference and a nearly exponential input atomic mode $\phi^{(out)}_1(z)$ (see
Fig.~\ref{psiout}a) the anti-Stokes field grows with $z$ sweeping the major part of the atomic polarization out of the
cell. At each $z$ along the pump beam, where the virtual beamsplitters mix atomic polarization and anti-Stokes light
the condition of suppressing the atomic output $\hat d(z,t)\sqrt{1-|\gamma|^2}=\gamma^*\hat c(z,t)$ is met, since the
anti-Stokes and atomic polarization both grow exponentially with $z$.
 For small group velocity difference $\Delta\beta'=-10\,$ps/mm, the input atomic polarization has a maximum inside the
cell and the anti-Stokes scattering process first shifts the atomic excitation into the field, but at some point the
process is partially reversed and the atomic polarization, albeit shifted towards the end of the cell, partially
remains within it. The atomic input mode shape is not proper to achieve suppression of the atomic output polarization
which occurred in the case of the of vanishing dispersion.
 Finally, for the largest group velocity difference $\Delta\beta'=-30\,$ps/mm, the initial atomic mode shape plays a minor
role, since the scattering processes in various parts of the cell are nearly independent from each other, because there
is no light connecting them. This is why the residual fraction of the atomic polarization can be smaller for larger
$\Delta\beta'$ as found in the numerical calculations and depicted in Fig.~\ref{Tvsg}.

We explain the oscillations of readout efficiency $1-\eta$ as a function of coupling $g'_0$ found in numerical
calculations and plotted in Fig.~\ref{Tvsg} in the following way. We investigate in detail the scattering in a small
part of the pumped region, shown in Fig.~\ref{fig:antiStokes}d. As soon as the atomic polarization reaches the pumped
region, it is converted into an anti-Stokes wave within a small period of time. Released light propagates along the
horizontal arrow, however soon it is converted back into atomic polarization. The excitation exchange continues until
the atomic polarization or light leaves the pumped region. Since the period of oscillations changes with the pumping
intensity, we observe in the end either excitation form alternately. In particular for some pumping intensities the
oscillations stop just when the excitations are stored in light, which corresponds to the maximum readout efficiency
$\eta$. This scenario is confirmed by the numerical simulations for high pumping intensity. It can be also obtained
from an analytical solution of \eeqref{eq:aprop} rewritten in the pump reference frame under assumption that the
quantum fields do not depend on $z$ in this reference frame.

\section{Conclusion}\label{sec:conclusions}
In summary, we have analyzed the Stokes and anti-Stokes scattering processes under transient pumping conditions and in
presence of group velocity difference between the interacting waves in the non-saturated regime. Using Bloch-Messiah
reduction \cite{BraunsteinBM} we confirm previous conjecture \cite{RaymerJMO04} that the Stokes scattering process can
be decomposed into multiple independent squeezers. Each of them entangles atoms and field, both in characteristic
input modes $\psi^{(in)}_n(t)$ and $\phi^{(in)}_n(z)$, into a pair of quantum-correlated states occupying
characteristic output modes $\psi^{(out)}_n(t)$ and $\phi^{(out)}_n(z)$. We compute those modes and their occupancies
$\avg{\hat n_n}$ for a few realistic cases. We find that with increasing group velocity difference between the pump and
Stokes waves, the scattering process involves an increasing number of modes which become comparably squeezed and occupied.
This is also reflected in the photon-count statistics we compute, which changes from single-mode thermal statistics in
the case of no group velocity difference, into multimode thermal statistics.

Next, we consider reading out the atomic polarization in the anti-Stokes scattering process. We show that in general
the readout can be decomposed into multiple independent beam-splitter like transformations. Each of them mixes the
statistics contained in a pair of characteristic input field $\Psi^{(in)}_n(t)$ and atomic $\Phi^{(in)}_n(z)$ modes,
producing two characteristic output modes field $\Psi^{(out)}_n(t)$ and atomic $\Phi^{(out)}_n(z)$. This fact stems
from the preservation of the total number of excitations during the anti-Stokes scattering.

However, when the anti-Stokes scattering is applied to reading out an atomic polarization in a particular mode, this
consideration can be much simplified. We show under what conditions the readout process is successful in translating the atomic
statistics into the field. It turns out that efficient readout is possible for wide range of group-velocity
differences, however usually the pumping intensity must be precisely controlled. Also it turns out that once the
readout is successful, it translates a given atomic mode into a predetermined field mode, and the latter is unsullied by
statistics of any other atomic modes.

Let us note that the equations governing Stokes and anti-Stokes scattering have closed-form anlytical solutions
\cite{Raymer1DTheory81} in the limit of vanishing dispersion.  In this limit modal decomposition can be obtained by a
single-step numerical diagonalization of the analytical field and atomic correlation functions \cite{RaymerJMO04}.

Finally let us remark, that our treatment applies to the case of writing a weak quantum signal into an atomic medium
\cite{KozehekinPRA00,DuanPRA02,NunnPriv05,EisamanPRL04,schori:057903}.


\begin{acknowledgements}
We acknowledge helpful discussions with Konrad Banaszek, Alex Lvovsky, Ian Walmsley, Joshua Nunn, Wenhai Ji, and Chunbai Wu, as well
as support from the Polish KBN grant numbers 2P03B 029 26 and 1 P03B 011 29, and the U.S. National Science Foundation,
grant numbers PHY0140370 and PHY0456974.
\end{acknowledgements}

\appendix
\section{Derivation of the nondegenerate reduction theorem}\label{app:BM}
The Bloch-Messiah reduction has been derived for a general case of a monolithic quantum system
\cite{BraunsteinBM}. The original theorem comprises derivation of a modal expansion of the system quantum field operator,
whose application reduces given output-input relations into a set of parallel single-mode squeezing transformations.

However, in the particular case of Raman scattering process two distinct types of operators exist --- those labelled by
space and those labelled by time. In this case, a fruitful extension of this theorem can be made. A natural division
into atomic and light subsystems which are separated before and after interaction is utilized for this. These
subsystems play symmetric roles in the process. This is a crucial fact we will employ to specify a particular form of
Bloch-Messiah reduction in the nondegenerate case. Below we will show how to introduce separate modal expansions for the
atomic and light subsystems and cast the input-output relations into parallel two-mode squeezing transformations.

We start our generalization of the Bloch-Messiah reduction theorem by first reducing the number of atomic and light
modes involved in the interaction to a finite number, so that we can use the original form of the theorem
\cite{BraunsteinBM}. This is easily done when one notes that the functions appearing in our considerations can be
approximated with any chosen accuracy on a discrete grid. Therefore we discretize the spatial-temporal region in which
the interaction occurs by introducing sets of points in time $\{t_n\}$ and space $\{z_n\}$ and restricting ourselves to
the rectangular grid spanned by this set. Then we can introduce the input and output vectors:
\begin{align}\label{eqa:discretization}
\vec{\hat{a}}_{\text{in},n}&= \hat a(0,t_n) &
\vec{\hat{a}}_{\text{out},n}&=\hat a(L,t_n) \nonumber \\
\vec{\hat{b}}_{\text{in},n}&= \hat b(z_n,-T) & \vec{\hat{b}}_{\text{out},n}&=\hat b(z_n,T)
\end{align}
the Green matrices:
\begin{align}
C_{a_{n,m}}&=C_a(t_n,t_m) &
C_{b_{n,m}}&=C_b(z_n,z_m) \nonumber \\
S_{a_{n,m}}&=S_a(t_n,z_m) & S_{b_{n,m}}&=S_b(z_n,t_m)
\end{align}
and the characteristic mode vectors:
\begin{align}
\vec\psi^{(in)}_{k,n} &=\psi^{(in)}_k(t_n) &
\vec\psi^{(out)}_{k,n}&=\psi^{(out)}_k(t_n) \nonumber\\
\vec\phi^{(in)}_{k,n} &=\phi^{(in)}_k(z_n) & \vec\phi^{(out)}_{k,n}&=\phi^{(out)}_k(z_n)
\end{align}
which allow for casting the equation \eqref{eq:io} into a matrix form:
\begin{eqnarray}\label{eqa:inout}
\begin{bmatrix} \vec{\hat{a}}_\text{out} \\ \vec{\hat{b}}_\text{out} \end{bmatrix}&=&
\underbrace{\begin{bmatrix} C_a & 0 \\ 0 & C_b \end{bmatrix}}_\mathbf{C}
\begin{bmatrix} \vec{\hat{a}}_\text{in} \\ \vec{\hat{b}}_\text{in} \end{bmatrix}
 +\underbrace{\begin{bmatrix} 0 & S_a \\ S_b & 0 \end{bmatrix}}_\mathbf{S}
\begin{bmatrix} \vec{\hat{a}}^\dagger_\text{in} \\ \vec{\hat{b}}^\dagger_\text{in}  \end{bmatrix}.
\end{eqnarray}
where the column vectors are concatenations of their components, and we defined block-matrices $\mathbf{C}$ and
$\mathbf{S}$. Since the atomic and light subsystems play a symmetric role in the process, the $\mathbf{C}$  commutes
with matrix $\mathbf{O}$ defined in the following way
\begin{equation}
\mathbf{O}=\begin{bmatrix} i & 0 \\ 0 & -i \end{bmatrix}
\end{equation}
while for $\mathbf{S}$ the following holds: $\mathbf{OS}=\mathbf{SO}^\dagger$.

The equation \eqref{eqa:inout} is a Bogoliubov transformation, thus the Bloch-Messiah reduction can be applied
\cite{BraunsteinBM} to $\mathbf{C}$ and $\mathbf{S}$ and we can express them in decomposed form:
\begin{eqnarray}\label{eqa:BM:C}
\mathbf{C}&=&\sum_k \cosh \zeta_k \,
\begin{bmatrix} \vec\psi^{(out)}_k, & \vec\phi^{(out)}_k \end{bmatrix}^\dagger \begin{bmatrix} \vec\psi^{(in)}_k, & \vec\phi^{(in)}_k \end{bmatrix} \\
\mathbf{S}&=&\sum_k \sinh \zeta_k \,
\begin{bmatrix} \vec\psi^{(out)}_k, & \vec\phi^{(out)}_k \end{bmatrix}^\dagger \begin{bmatrix} \vec\psi^{(in)}_k, & \vec\phi^{(in)}_k \end{bmatrix}^*
\label{eqa:BM:S}
\end{eqnarray}
where $[\vec\psi^{(out)}_k, \vec\phi^{(out)}_k ]$ and $[\vec\psi^{(in)}_k, \vec\phi^{(in)}_k]$ are left and right
eigenvectors (output and input eigenmodes), defined as concatenation of respective component vectors,
$\zeta_n$ are real numbers. Here $\dagger$ denotes transposition and complex conjugation. By definition of singular value
decomposition \cite{SVDReview} the singular vectors obey the following relations:
\begin{eqnarray}
\mathbf{C}\begin{bmatrix} \vec\psi^{(in)}_k, & \vec\phi^{(in)}_k \end{bmatrix}^\dagger&=&
\cosh \zeta_k  \begin{bmatrix} \vec\psi^{(out)}_k, & \vec\phi^{(out)}_k \end{bmatrix}^\dagger \\
\mathbf{C}^\dagger\begin{bmatrix} \vec\psi^{(out)}_k, & \vec\phi^{(out)}_k \end{bmatrix}^\dagger&=& \cosh \zeta_k
\begin{bmatrix} \vec\psi^{(in)}_k, & \vec\phi^{(in)}_k \end{bmatrix}^\dagger
\end{eqnarray}
multiplying both equations by $\mathbf{O}$ from the left and using $\mathbf{CO}=\mathbf{OC}$ shows, that if
$[\vec\psi^{(out)}_k, \vec\phi^{(out)}_k ]$ and $[\vec\psi^{(in)}_k, \vec\phi^{(in)}_k]$ are left and right
eigenvectors, then $[i\vec\psi^{(out)}_k, -i\vec\phi^{(out)}_k ]$ and $[i\vec\psi^{(in)}_k, -i\vec\phi^{(in)}_k]$ also
satisfy the above relations, and thus by definition of SVD are left and right eigenvectors of $\mathbf{C}$. Therefore
in the summation \eqref{eqa:BM:C} we can group those vectors together, obtaining:
\begin{align}
\mathbf{C}=&\sum_k \cosh \zeta_k \, \Biggl(
\begin{bmatrix} \vec\psi^{(out)}_k, & \vec\phi^{(out)}_k \end{bmatrix}^\dagger \begin{bmatrix} \vec\psi^{(in)}_k, & \vec\phi^{(in)}_k \end{bmatrix} \nonumber\\
&+\begin{bmatrix} i\vec\psi^{(out)}_k, & -i\vec\phi^{(out)}_k \end{bmatrix}^\dagger \begin{bmatrix} i\vec\psi^{(in)}_k, & -i\vec\phi^{(in)}_k \end{bmatrix}\Biggr) \nonumber\\
=&\sum_k \cosh \zeta_k\begin{bmatrix} \vec\psi^{(out)\dagger}_k \vec\psi^{(in)}_k & 0 \\ 0 & \vec\phi^{(out)\dagger}_k
\vec\phi^{(in)}_k \end{bmatrix}
\end{align}
Which shows that in each term of the singular value decomposition of $\mathbf{C}$ the off-diagonal elements vanish.
Applying analogous transformation to the summation \eqref{eqa:BM:S} it can be confirmed that the same property holds for $\mathbf{S}$.
These statements are the content of our extension of Bloch-Messiah reduction. They allow for expressing the
blocks of $\mathbf{C}$ as in \eeqref{eq:CSdecompCa} and \eqref{eq:CSdecompCb} in the text, and blocks of
$\mathbf{S}$ as in \eeqref{eq:CSdecompSa} and \eqref{eq:CSdecompSb}.

\section{Proof of decomposition into beamsplitter transformations}\label{app:BS}
Analogously to Appendix~\ref{app:BM} we begin by discretizing the annihilation operators
\begin{align}\label{eqb:discretization}
\vec{\hat{c}}_{\text{in},n}&= \hat c(0,t_n) &
\vec{\hat{c}}_{\text{out},n}&=\hat c(L,t_n) \nonumber \\
\vec{\hat{d}}_{\text{in},n}&= \hat d(z_n,-T) & \vec{\hat{d}}_{\text{out},n}&=\hat d(z_n,T)
\end{align}
the Green matrices:
\begin{align}
C_{c_{n,m}}&=C_c(t_n,t_m) &
C_{d_{n,m}}&=C_d(z_n,z_m) \nonumber \\
S_{c_{n,m}}&=S_c(t_n,z_m) & S_{d_{n,m}}&=S_d(z_n,t_m)
\end{align}
and the characteristic mode vectors:
\begin{align}
\vec\Psi^{(in)}_{k,n} &=\Psi^{(in)}_k(t_n) &
\vec\Psi^{(out)}_{k,n}&=\Psi^{(out)}_k(t_n) \nonumber\\
\vec\Phi^{(in)}_{k,n} &=\Phi^{(in)}_k(z_n) & \vec\Phi^{(out)}_{k,n}&=\Phi^{(out)}_k(z_n)
\end{align}
Next, let us rewrite \eeqref{eq:aio} in a matrix form:
\begin{eqnarray}\label{eqb:inout}
\begin{bmatrix} \vec{\hat c}_\text{out} \\ \vec{\hat d}_\text{out} \end{bmatrix}&=&
\underbrace{\begin{bmatrix} C_c & S_c\\ -S_d & C_d \end{bmatrix}}_\mathcal{U}
\begin{bmatrix} \vec{\hat c}_\text{in} \\ \vec{\hat d}_\text{in} \end{bmatrix}
\end{eqnarray}
For the energy to be conserved in the anti-Stokes scattering process, the underbraced matrix $\mathcal{U}$ must be
unitary
\begin{equation}\label{eqb:unitary}
\begin{bmatrix} C_c & S_c\\ -S_d & C_d \end{bmatrix}
\begin{bmatrix} C_c^\dagger & -S_d^\dagger\\ S_c^\dagger & C_d\dagger \end{bmatrix}=
\begin{bmatrix} C_c^\dagger & -S_d^\dagger\\ S_c^\dagger & C_d\dagger \end{bmatrix}
\begin{bmatrix} C_c & S_c\\ -S_d & C_d \end{bmatrix}=
\begin{bmatrix} \mathbf{1} & 0 \\ 0 & \mathbf{1}\end{bmatrix}
\end{equation}
therefore eight equations constrain the Green matrices. Let us first focus on the diagonal parts of the product. The
condition $C_cC_c^\dagger+S_cS_c^\dagger=1$ means that matrices $C_cC_c^\dagger$ and $S_cS_c^\dagger$ commute,
therefore  $C_cC_c^\dagger$ and $S_cS_c^\dagger$ have the same eigenvectors. This in turn imposes the condition that
the left singular vectors of $C_c$ and $S_c$ are the same. By analyzing the remaining equations contained in the
diagonal portion of \eeqref{eqb:unitary} we conclude that the singular vectors are shared between Green functions. Thus
we can specify their singular value decompositions:
\begin{align}\label{eqb:BCSdecomp}
C_c=&\sum_n \alpha_n \vec\Psi^{(out)*}_n\, \vec\Psi^{(in)T}_n &
C_d=&\sum_n \beta_n  \vec\Phi^{(out)*}_n\, \vec\Phi^{(in)T}_n \nonumber \\
S_c=&\sum_n \gamma_n \vec\Psi^{(out)*}_n\, \vec\Phi^{(in)T}_n & S_d=&\sum_n \delta_n \vec\Phi^{(out)*}_n\,
\vec\Psi^{(in)T}_n.
\end{align}
Let us note that all the singular values can be always made real, by transferring the phase factor into the associated
singular vectors. Inserting these Green matrices in the above form into \eeqref{eqb:unitary} yields, that for every $n$
the matrix built from the eigenvalues
\begin{equation}
\mathbf{U}_n= \begin{bmatrix} \alpha_n & \gamma_n\\ -\delta_n & \beta_n \end{bmatrix}
\end{equation}
must be unitary. But any real unitary 2x2 matrix has the form
\begin{equation}
\begin{bmatrix} \alpha_n & \gamma_n\\ -\delta_n & \beta_n \end{bmatrix}=
\begin{bmatrix} \cos\theta_n & \sin\theta_n\\ -\sin\theta_n & \cos\theta_n \end{bmatrix}
\end{equation}
where $\theta_n$ is a real angle. Inserting the singular values in the above form into \eeqref{eqb:BCSdecomp} and
replacing $\sin\theta_n=\sqrt{\eta_n}$ we get \eeqref{eq:BCSdecomp}.

\end{document}